\documentclass[grl]{agu2001}
 \usepackage{graphicx}
\graphicspath{{figures/}}
\authorrunninghead{GERMAIN ET AL.}
\titlerunninghead{GNSS-R speculometry  from low aircraft}
\authoraddr{G. Ruffini, Research Department, Starlab, C. de l'Observatori, s/n, 08035 Barcelona, Spain. (giulio.ruffini@starlab.es)}
\begin{document}

%% ------------------------------------------------------------------------ %%
%
%  TITLE
%
%% ------------------------------------------------------------------------ %%

\title{The Eddy Experiment: GNSS-R speculometry for directional sea-roughness retrieval from low  altitude aircraft}
%
% e.g., \title{Terrestrial Ring Current:
% Origin, Formation and Decay $\alpha\beta\Gamma\Delta$}

%% ------------------------------------------------------------------------ %%
%
%  AUTHORS AND AFFILIATIONS
%
%% ------------------------------------------------------------------------ %%

% Method 1 (for all journals, except Reviews of Geophysics, which
% should use method 3):
% For three or fewer author/affiliation blocks, use \author{} and \affil{}

%\author{G. Ruffini, F. Soulat, M. Caparrini and O. Germain}
%\affil{ Starlab, C. de l'Observatori Fabra s/n, 08035 Barcelona, Spain, http://starlab.es}

% ---------------
% Method 2 (for all journals, except Reviews of Geophysics, which
% should use method 3): For more than three author/affiliation blocks,
% use \author{\altaffilmark{}} and \altaffiltext{}
% \altaffilmark will produce footnote; matching altaffiltext
% will appear at bottom of page. May use \\ to start a new line.

\authors{O. Germain, G. Ruffini, F. Soulat, M. Caparrini,  \altaffilmark{1}
B. Chapron \altaffilmark{2}
 and P. Silvestrin \altaffilmark{3}}

\altaffiltext{1}
{Starlab, C. de l'Observatori Fabra s/n, 08035 Barcelona, Spain, http://starlab.es}
\altaffiltext{2}
{Ifremer, Technop\^ole de Brest-Iroise BP 70, 2920 Plouzan\'e, France, http://ifremer.fr}
\altaffiltext{3}{ESA/ESTEC, Keplerlaan 1, 2200 Noordwijk, The Netherlands,  http://esa.int}
%
% \altaffiltext{3}{Department of Space Sciences, University of Michigan,
% Ann Arbor, Michigan, USA.}
%
% \altaffiltext{4}{Desert Research Institute, Division of Hydrologic Sciences,
% Reno, Nevada, USA.}

%---------------
% Method 3 (for Reviews of Geophysics only): Reviewauthors is a table with three
% columns. You must supply the ''&'' between each author/affiliation. If you have
% more than three authors, start a new table line with /cr

% e.g.,
% \begin{reviewauthors}
% R. C. Bales\\
% Department of Hydrology and\\ Water Resources\\
% University of Arizona\\
% Tucson, Arizona, USA
% &
% E. Mosley-Thompson\\
% Department of Geography\\
% Ohio State University\\
% Columbus, Ohio, USA
% &
% J. R. McConnell\\
% Desert Research Institute\\
% Reno, Nevada, USA
% \end{reviewauthors}

%% ------------------------------------------------------------------------ %%
%
%  ABSTRACT
%
%% ------------------------------------------------------------------------ %%

% Do NOT include any \begin...\end commands within
% the body of the abstract.

\begin{abstract}
We report on the retrieval of directional sea surface roughness, in terms of its full directional mean square slope (including  direction and isotropy), from  Global Navigation Satellite System Reflections (GNSS-R) Delay-Doppler-Map (DDM)  data collected during an experimental flight at 1 km altitude. This study emphasizes the utilization of the entire DDM to more precisely infer ocean roughness directional parameters. In particular, we argue that the DDM exhibits the impact of  both roughness and scatterer velocity. Obtained estimates are analyzed and compared to co-located Jason-1 measurements, ECMWF numerical weather model outputs and optical data. 
\end{abstract}

%% ------------------------------------------------------------------------ %%
%
%  TEXT
%
%% ------------------------------------------------------------------------ %%

% The body of the article must start with a \begin{article} command,
% and an \end{article} command must follow the references section.
% Otherwise, the text will not print at the appropriate column width.
%

\begin{article}

\section{Introduction}

\begin{figure} [b!]
   \hspace{-0.4cm}      \includegraphics[width=8cm]{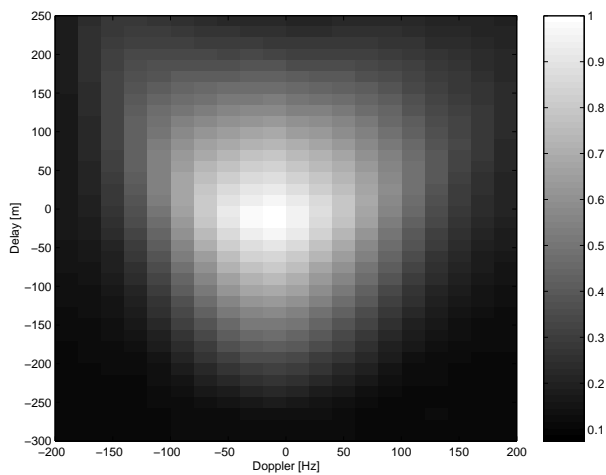} 
    \caption{Example of GPS-R Delay-Doppler Map.}
  \label{ddm_example}
\end{figure}

\begin{figure}[t!]
 \begin{tabular}{c}
\includegraphics[width=7cm]{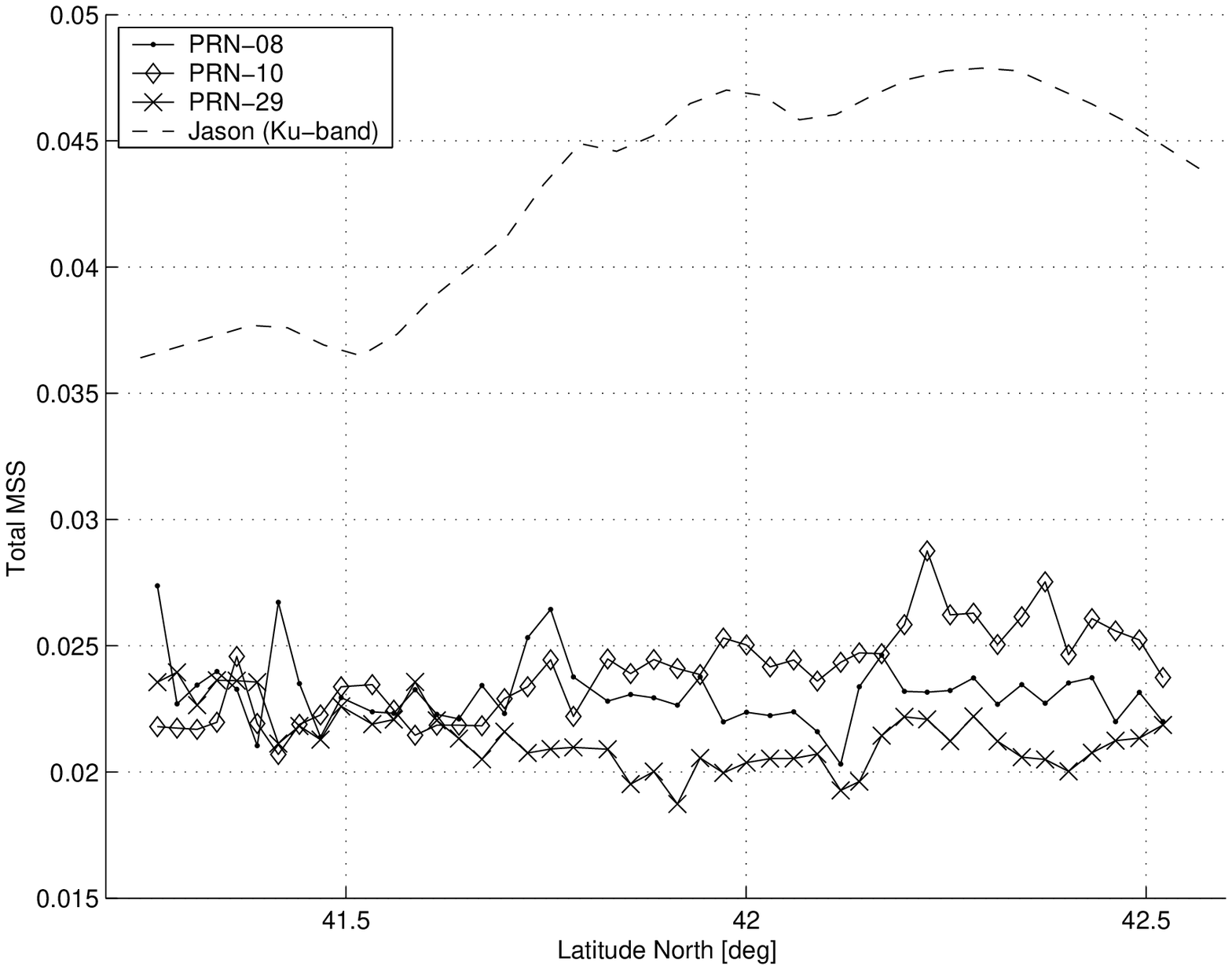} \\
\includegraphics[width=7cm]{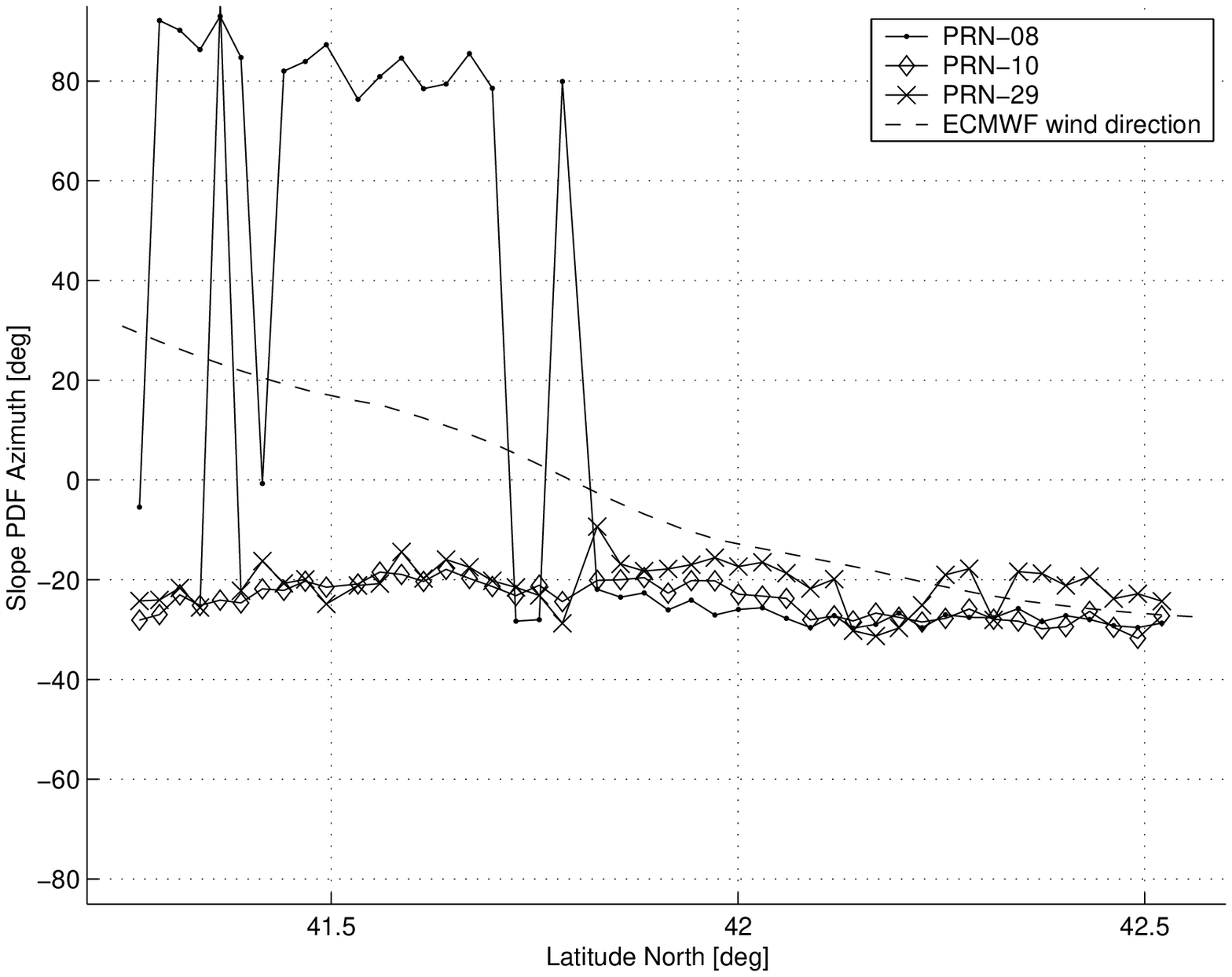} \\
\includegraphics[width=7cm]{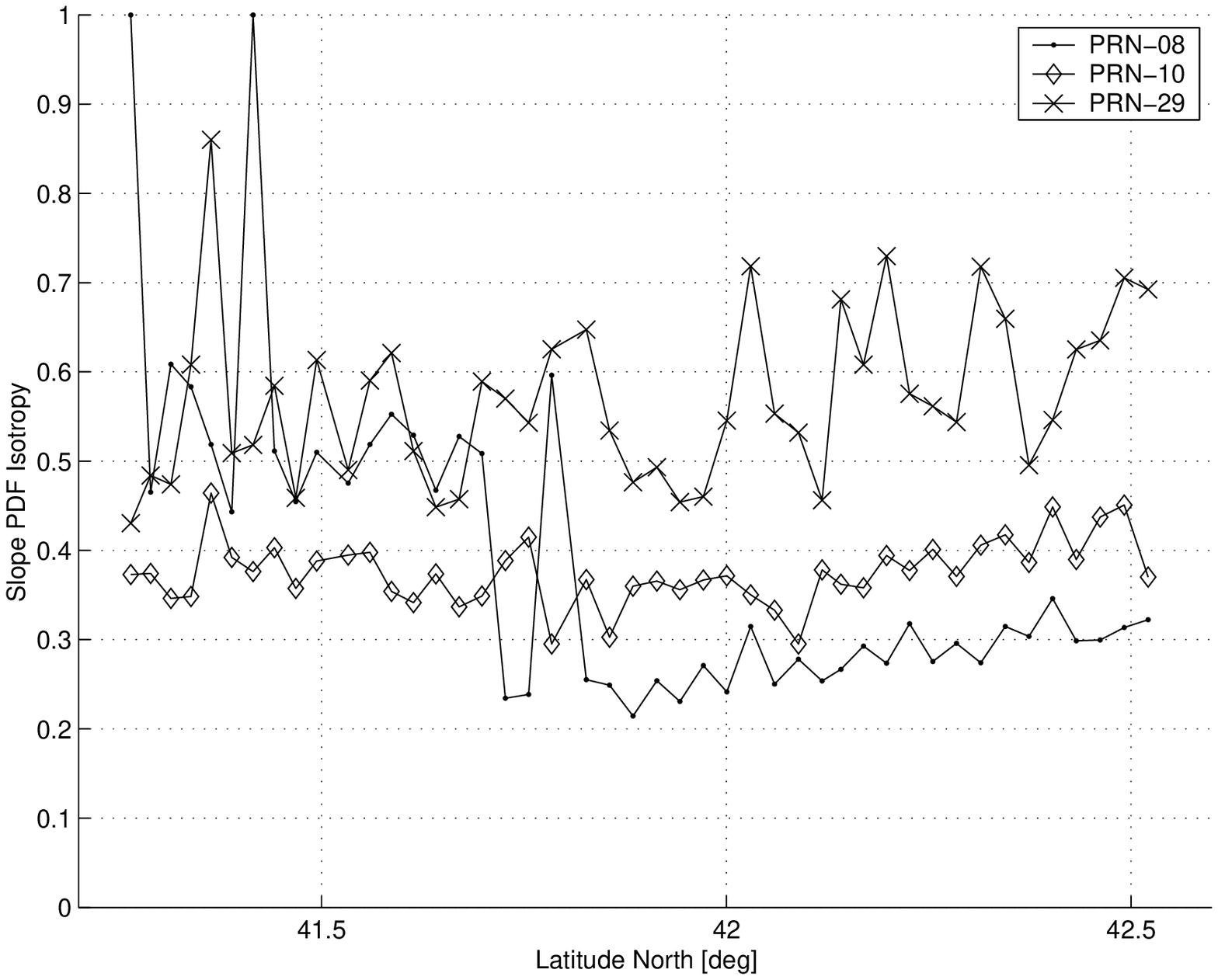}
\end{tabular}
    \caption{DMSS$_\lambda$ estimation with the DDM least-square inversion approach, along the descending (North to South) track---MSS top, SPA middle and SPI bottom.
The total MSS (Ku-Band) measured by Jason-1 and the ECMWF wind direction are also shown for comparison.}
  \label{DMSS}
\end{figure}

%\medskip
Several GNSS constellations and augmentation systems are presently operational or under development, including the  pioneering US Global Positioning System (GPS) and the forthcoming European system, Galileo. These all-weather, long-term, stable and precise L-band signals can be used for bistatic remote sensing of the ocean surface and beyond, an emerging concept known as GNSS-R.

Among several applications, two have been emphasized by the community:
sea-surface altimetry (see \markcite{{\it Ruffini et al.} [2004]} and references therein) and sea-surface  ``speculometry'' (a term discussed below),  related to the statistical properties of sea surface gravity wave slopes. Although this paper addresses the latter, 
%GNSS-R holds the potential to  provide the unprecedented spatio-temporal samplings needed for mesoscale monitoring of ocean circulation.
%Global, mesoscale measurements of the sea-surface roughness would be of great value to GCOS/GOOS (Global Climate/Ocean Observing System), especially if they are collocated with sea-surface height measurements.
we note the intrinsic capability of GNSS-R for providing long term co-located  measurements of  both surface roughness and sea level  with  high spatial and temporal resolution.  As recently demonstrated with scatterometer measurements (\markcite{{\it Chelton et al.} [2004]}),  such a capability  would help to better quantify the relationship of velocities in the upper ocean  (driven by wind stress forcing) with surface height dynamics.

Thanks to its passive multistatic character, GNSS-R clearly holds the potential to  provide an unprecedented spatio-temporal sampling of the ocean surface. The expected high spatial and temporal measurements can certainly serve operational applications. For instance, and to follow successful scatterometer measurements, GNSS-R can complement ocean winds and wave models. Being rain immune, such new data could  help quantify atmosphere-ocean coupling, including momentum and energy fluxes under extreme conditions for hurricane modeling. Many other scientific applications can be cited, such as sea surface breaking/whitecapping and gas exchange global characterization,  important ingredients for understanding the ocean's biogeochemical response to, and its influence on, climate change. Indeed,
a very promising approach to quantify CO$_2$ flux  is to better assess the surface fractional area, which is readily measurable from surface slope measurements  
%Recent studies at NASA/Wallops (FEDS.98, the Flux Exchange Dynamics Study) indicate that surface roughness (through its relation to Fractional Area) is a much better quantifier of the air sea gas flux than the traditionally used wind speed
 (\markcite{{\it Watson et al.} [1999]}). In that case,  GNSS-R can provide  a more direct measurement to extract gravity surface slope statistical properties and to quantify the role of the ocean in taking up increases of CO$_2$.

In addition, L-band sea surface roughness data can be used to support L-band radiometric missions, such as SMOS (Soil Moisture and Ocean Salinity) and AQUARIUS, to both quantify and efficiently separate roughness and salinity contributions to L-band radiometric brightness measurements. %GNSS-R measurements will help interpret and complement radiometric  salinity measurements as well as suggest synergic uses of both types of missions.

%%%%%%%%%%%5

Inferring sea roughness from GNSS-R data requires (i) a parametric description of the sea surface, (ii) an electromagnetic and instrument model for sea-surface scattering at L-band and (iii) the choice of a GNSS-R data product to be inverted. 
There is quite an agreement on the two first aspects in the literature.
It has been recognized that the scattering of GNSS signals can be  approximated by  an {\it effective} Geometric  Optics model, where the fundamental physical process is the scattering from facet-like surface elements. This is the reason for the use of the term ``speculometry'' here, which stems  from the Latin word for mirror, {\it speculo}: the detected GNSS-R return is dominated by the statistics of facet slopes and their curvatures at scales larger than the electromagnetic wavelength ($\lambda$). 

%The statistical distribution of facet slopes is described by  the bi-dimensional sea-surface slope probability density function (PDF). 
Under a Gaussian assumption,  three parameters fully define the detected L-band sea surface slope probability distribution (PDF). These parameters are encapsulated by  the directional mean square slope, DMSS$_\lambda$, a symmetric tensor which results from  the integration of the ocean energy spectrum at wavelengths larger than $\lambda$, and which characterizes the ellipsoidal shape of the slope PDF.
%: scale (total MSS), direction (Slope PDF azimuth) and isotropy (Slope PDF isotropy).
%%%%%%55

%Sea-surface near-specular scattering of GNSS signals can indeed be  fairly well described by a Geometric Optics mechanism (GO), where the fundamental physical process is the scattering from mirror-like surface elements (hence the term ``speculometry'', from {\it speculo}, the Latin word for mirror). Therefore, the most important feature of the sea surface is the statistics of facet slopes at about the same scale as the electromagnetic wavelength ($\lambda$), as described by the bi-dimensional slope probability density function (PDF). 
It is important to note that DMSS$_\lambda$  has  rarely been emphasized as the geophysical parameter of interest in the literature. Instead, most authors link sea roughness to the near surface wind vector, which is thought to be more useful for oceanographic and meteorological users. Unfortunately, this is somewhat misleading, as the relationship between surface wind and roughness is not one-to-one  and  requires an additional modeling layer. 
%For instance, a wind-driven sea spectra is not suitable for inferring sea surface DMSS$_\lambda$ when swell is present or the sea not fully developed. 
The connection between DMSS$_\lambda$ and wind is affected by other factors (e.g., swell, fetch and degree of maturity), as is well known in the altimeter community  (see \markcite{{\it Gourrion et al.} [2002]}).

Moreover,  the product traditionally used for inversion in GNSS-R speculometry is the  1D delay waveform of the reflected signal amplitude, from which  the wind speed is inferred assuming an isotropic slope PDF (see, e.g.,
 \markcite{{\it Garrison et al.} [1998, 2002]}
\markcite{{\it Komjathy et al.} [2000]}, or
%\markcite{{\it Garrison et al.} [1998, 2002]}, or
\markcite{{\it Cardellach et al.} [2003]}).
Attempts have also been made to estimate the wind direction by fixing the PDF isotropy to some theoretical value (around 0.7) and using at least two satellites reflections with different azimuths (see, e.g., 
\markcite{{\it Zuffada et al.} [2000]},
\markcite{{\it Armatys et al.} [2000]} and \markcite{{\it Garrison et al.} [2003]}).

Here, we will work with a product of higher information content:   the 2D DDM of the reflected signal amplitude.
As proposed in \markcite{{\it Ruffini et al.} [2000]},
the provision of an extra dimension opens the possibility of performing a robust  estimation of all the DMSS$_\lambda$ parameters through the direct fitting/estimation of the entire DDM. In  \markcite{{\it Elfouhaily et al.} [2002]}, a first order approximation for  inversion was proposed.

The full DDM-inversion technique proposed in \markcite{{\it Ruffini et al.} [2000]} is used for the first time here to analyze GNSS-R data collected during the Eddy Experiment. This campaign and the  altimetric data analysis is reported elsewhere (\markcite{{\it Ruffini et al.} [2004]} and \markcite{{\it Soulat} [2003]}), and is only briefly described in Section~2.
%(where the analysis of  optical data was also carried out).   
%The primary goal of the paper is to investigate the full exploitation of the bi-dimensional GNSS-R DDM  product to infer the set of three DMSS$_\lambda$ parameters.  
%The driver of the study has been the exhaustive exploitation of the information contained in the DDM product.  
The retrieval methodology relies on a least-squares fit  of the speculometric model, as discussed in the third section. In the fourth section, results are compared to ancillary data (Jason-1 radar altimeter, ECMWF numerical weather model, optical data). Finally, another important outcome of the exhaustive exploitation of the information contained in the DDM product is related to the expected mean sea surface motion. We evidence that a small part of the Doppler spread can be attributed to the mean ``scatterer velocity'', i.e., the rapid motion of the $\lambda$ (or larger) sized  sea-surface facets that contribute the most to the detected signals.

\section{Data collection and pre-processing}

The data set (i.e., the recorded direct and reflected GPS signals together with the aircraft kinematic data) was gathered during an  airborne campaign carried out in September~2002. The aircraft overflew the Mediterranean Sea, off the coast of Catalonia (Spain), northwards from  the city of Barcelona for about 150~km at 1000 m altitude and 45-75 m/s speed.
The area is crossed by the ground track \#187 of the Jason-1 radar altimeter, which the aircraft  overflew during the satellite overpass for precise comparison.
The track was  overflown twice: the first time during the ascending pass (from South to North) at low speed (45-60 m/s) and  the second  during the descending pass (from North to South) at a faster speed (65-75 m/s) due to  wind. The time shift between the two passes over a same point on the track  ranged from 45 min to 2h 15 min. During the ascending pass, PRNs 08, 10 and 24 were visible with elevations spanning  30$^o$ to 85$^o$ while PRNs 08, 10 and 29 were visible during the descending track with elevations between 40$^o$ and 75$^o$.
The configuration of this test flight was not optimized for speculometry: from such low altitude, the sea-surface reflective area is essentially limited by the PRN C/A code, and the glistening zone is coarsely Delay-Doppler mapped.

The raw GPS signals were acquired with a modified TurboRogue receiver, sampled at 20.456 MHz and pre-processed  with a dedicated software composed of two sub-units fed with the direct and reflected signals. 
%The software direct-signal unit uses standard algorithms to track the correlation peak of the signal, both in time and frequency. The reflected-signal unit performs correlations blindly, with  time delay and frequency settings which depend on those of the  direct-signal unit. C
Correlations were computed at 81 delay lags while the Doppler dimension spanned -200 to 200~Hz with a step of 20~Hz. The coherent/incoherent integration times were respectively set to 20~ms and 10~s, meaning that the averaged DDM were produced at the rate of 0.1~Hz after summation of 500 incoherent looks (see Figure~\ref{ddm_example} for a sample DDM).

%%%%%%%%%%%%%%%%%%%%%%%%%%%%%%%%%%%%
%%%%%%%%%%%%%
\section{Speculometric model and DDM inversion}
%The GNSS-R scattering model proposed by~\markcite{{\it Zavorotny et al.} [2000]} is, to date, the reference model for the community. 
Following \markcite{{\it Zavorotny et al.} [2000]}, the link between the DDM mean power at delay-Doppler $P(\tau,f)$ and the effective L-band sea-surface slope PDF, ${\cal P}(s_x,s_y)$, is given by
\begin{eqnarray}
P(\tau,f) & = &
\int  dx dy \,
\frac{G_r}{R_t^2 R_r^2} \cdot
\frac{q^4}{q_z^4} \cdot
~{\cal P} \left( \frac{-q_x}{q_z},\frac{-q_y}{q_z} \right) 
 \cdot \nonumber \\
&& \; \: \; \:
\chi^2 \left[
\tau_m(x,y)-\tau_c-\tau,f_m(x,y)-f_c-f
\right] , 
\label{radar_eq}
\end{eqnarray}
where
$G_r$ is the receiver antenna pattern,
$R_t$ and $R_r$ the distances from generic point on sea-surface to transmitter and receiver,
$(q_x,q_y,q_z)$ the scattering vector,
$\chi$  the Woodward Ambiguity Function (WAF), 
$\tau_m(x,y)$ and $f_m(x,y)$  the delay-Doppler coordinates on the sea-surface and
$(\tau_c,f_c)$  the delay/Doppler offset of the geometric specular-point with respect to the direct signal (the DDM ``center'').
Accounting for the receiver mean thermal noise $P_N$ 
%(which can be measured in the early delay lags of the DDM) 
and including a scaling parameter $\alpha$, the mean amplitude of the DDM can be written as $
A(\tau,f) = \sqrt{\alpha \, P(\tau,f) + P_N}$.
As discussed above, ${\cal P}$ is described by the DMSS$_\lambda$ parameter set, which defines an elliptic quadratic form in the 2D space of facet slopes. Mean-square slopes along major and minor principal axes are often called MSS up-wind ($mss_u$) and MSS cross-wind ($mss_c$) respectively. In the following, we will refer to the Total MSS (MSS$_{tot}$=$2 \sqrt{mss_u . mss_c}$, proportional to the ellipse area and directly related to nadir $\sigma^o$), the Slope PDF azimuth (SPA, the direction of semi-major axis with respect to North) and the Slope PDF Isotropy (SPI, equal to $mss_c / mss_u$).

The inversion was performed through minimization of the mean square difference between model and data DDMs. Numerical optimization was carried out by a steepest-slope-descent algorithm with a Levenberg-Marquardt type adjustment. The main difficulty stemmed from the presence of several nuisance parameters in the forward model (mainly $\tau_c$ and $f_c$ but also $\alpha$).
The DDM centers were affected by the aircraft trajectory (altitude and vertical velocity) to first order but also by geophysical parameters (such as sea level). They needed to be accurately known in order to estimate DMSS$_\lambda$. For this reason, the DMSS$_\lambda$ and nuisance parameters were jointly estimated in an iterative manner.%: nuisance parameters (as a first step) and DMSS$_\lambda$ (as a second step) were successively searched, repeating this two-step sequence until convergence.
%The process was initialized with the following values:
%$\tau_c = f_c = 0$, $\alpha=1$, MSS$_{tot}$=0.025, SPA=0, SPI=0.65.

\section{Results and analysis}

%\subsection{DMSS$_\lambda$ estimation}

The values of DMSS$_\lambda$ estimated along the descending track of the flight are shown on Figure~\ref{DMSS}. The top plot illustrates the variations of Total MSS. The inter-PRN consistency is reasonable in the southern part of the track but worsens slightly in the northern part. For comparison, the total MSS in Ku-band was derived from the Jason-1 $\sigma^0$ co-located measurements at 1~Hz sampling (7~km) and 20~km resolution. The Jason-1 MSS was obtained through the simple relationship MSS$=\kappa / \sigma^0$, $\kappa$ being the effective (empirical) Fresnel coefficient, here set to 0.45. As expected, we observed that the level and dynamic of MSS decreased with longer wavelength (from 2~cm in Ku-band to 19~cm in L-band). The lower dynamic of L-band MSS impeded any clear trend comparison, although the measurements of PRN-10 seem in good agreement with Jason-1. The Jason-1 wind speed, derived from both the Ku-band $\sigma^0$ and the significant wave height (of about 2~m), ranged from 9 to 13 m/s along the track. Translating this wind speed into L-band MSS through the spectrum of \markcite{{\it Elfouhaily et al.} [1997]} yields values between 0.0220 and 0.0255,  in-line with  GNSS-R results. However, we must emphasize that the assumption of a wind-driven spectrum was not really warranted during the campaign. 
%In particular, the presence of swell definitely had a significant impact.

SPA estimation results are presented on the middle plot. The inter-PRN consistency is here very satisfying, and the apparent discrepancy in the southern part of the track can be explained as a degenerate solution of the estimation problem. Indeed,
the inversion of DDM for the SPA is degenerate in  at least two cases: when the transmitter is at zenith or when the receiver moves towards the transmitter. In these scenarios, the Delay-Doppler lines mapping the glistening zone are fully symmetric around the receiver direction and it becomes impossible to distinguish a slope PDF from its mirror image about the receiver direction. This effect is clearly observed here where the two found SPA  (-20$^o$ and 80$^o$) are indeed symmetric around the aircraft heading direction (30$^o$). In this part of the track, the azimuth of PRN-08 is about 50$^o$, almost aligned with the aircraft heading direction. In the northern part of the track, the estimated SPA matches very well with the wind direction provided by ECMWF. In the southern part, the mismatch reaches up to 50$^o$. However, we underline that surface wind is not the only element driving SPA and that swell is  likely to have contributed. 
%Besides, the accuracy of ECMWF wind direction is estimated to  20$^o$.

Finally, the bottom plot shows the SPI variations along the track. The reflected signals are strongly directional. The wind-driven spectrum of \markcite{{\it Elfouhaily et al.} [1997]} for a mature sea predicts a SPI value around 0.65, largely insensitive to wind speed. The apparent significant departure from this reference value is a probable signature of an under-developed sea with the presence of swell. However, the relatively poor consistency among PRN remains an issue to clarify: further work is needed to validate the accuracy of these SPI estimates and to better understand the potential information (and possible applicability constraints) of this product.

%\subsection{Scatterer velocity}

As a second and new outcome of this analysis, we now discuss the signature of ``scatterer velocity'' in the data, i.e., the signature of fastly moving sea-surface facets with size (curvature) larger than $\sim$20~cm. Such a signature can be detected when comparing the total MSS along the ascending and descending tracks, as estimated by the least-squares approach: a drastic discrepancy (up to 33\%) was observed for two passes shifted by less than hour over the same track point.
\begin{figure} [b!]
   \hspace{-0.4cm}     \includegraphics[width=8cm]{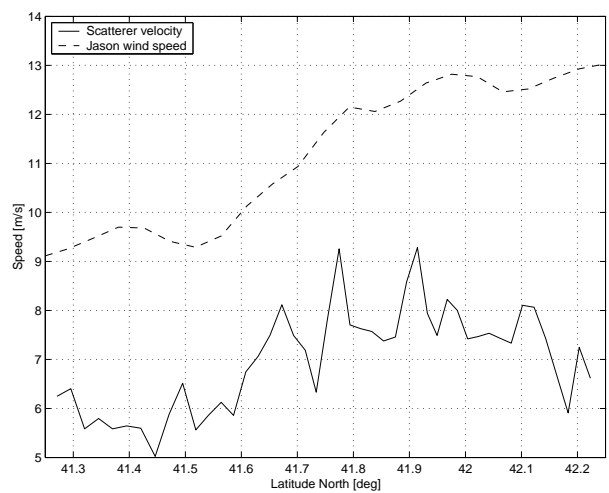} 
    \caption{Average scatterer velocity obtained when assuming a perfect match of ascending/descending MSS and the first order MSS model. It  correlates fairly well with wind speed and the observed swell from optical data.}
  \label{mean_scatterer_velocity}
\end{figure}

Multipath effects could conceivably lead to a Doppler width modulation. However,  some azimuth dependence should have been observed and was not.
Another possible cause  could be a changing aircraft attitude between ascending and descending tracks, but the aircraft roll and pitch values were checked to be nominal along both tracks. Moreover, a change in yaw would slightly impact the antenna pattern ground projection but would translate only into a Doppler bandwidth cut and never  a broadening. 
We concluded that the most likely explanation was of geophysical origin: the Doppler spectral width had been modulated  by the overall sea surface dynamics as detected from the mean specular facet motion. While the inversion approach assumed a still surface,  the relative velocity between receiver and scatterer should be taken into account for proper DDM inversion. At high receiver velocities this assumption is fine because the scatterer velocity impact will not be significant. At low speeds, however,  scatterer velocity becomes relevant.
This analysis is consistent with the fact that the MSS estimated in the ascending track (slower aircraft speed, from 45 to 60 m/s) showed abnormal high values compared to the ones estimated during the ascending track (faster speed, from 65 to 75 m/s).

In order to test this idea in a  simple manner, we used the results in \markcite{{\it Elfouhaily et al.} [2002]}, where it was demonstrated that 
a  first order relationship exists between the moments of the DDM and the full set of DMSS$_\lambda$ if the impact of the bistatic WAF and antenna gain are neglected: MSS=$\lambda^2 B^2/ (2 V^2 \sin^2 \epsilon)$,
where $\epsilon$ is the transmitter elevation, $V$ the receiver speed and $B$ the DDM Doppler bandwidth.
Assuming that the  MSS did not vary significantly between the times of the ascending and descending passes and considering the relative velocity between sea-surface scatterers and aircraft, i.e., $V\pm v_s$ for the descending/ascending passes respectively, we can roughly solve for the scatterer velocity $v_s$ (for simplicity assumed here parallel to the ground track).
Applying this scheme to all possible pairs of  measurements and averaging, the plot of Figure~\ref{mean_scatterer_velocity} results. The apparent mean scatterer velocity variations correlate with the JASON-1 wind speed variations.  Moreover,  from the dispersion relation an average scatterer speed of 8 m/s corresponds to waves of about 45~m wavelength. The detected motion thus seems to be associated with the longer and more coherent wave components, consistent with  optical observations of the swell vector (wavelength and direction, see \markcite{{\it Soulat} [2003]}), revealing the presence of a northerly (almost aligned with the flight track) generated 48~m wave system. 
As hypothesized, the quasi-specular facets contributing the most to the bistatic reflected signals may, on average,  be assumed to almost coherently travel with the dominating longer wave peak component.

\section{Conclusion}

We have reported the inversion of GNSS-R signals using, for the first time, the full Delay-Doppler Maps for the retrieval of the sea-surface directional mean square slope, DMSS$_{\lambda}$: the estimates show good inter-PRN consistency (except for the measured anisotropy SPI) and fair agreement with other sources of data. 

The use of the full DDM further has helped to reveal a geophysical signature in GNSS-R associated with the mean sea surface scatterer velocity. Under quasi-specular conditions, sea surface scatterers are mostly associated  to small slopes corresponding to longer waves with  velocities that can reach 5-10 m/s, impacting significantly the Doppler bandwidth of slow-moving receivers (e.g., airborne or ground-based, \markcite{{\it Soulat} [2004]}). The detection of such a geophysical signature opens new opportunities for GNSS-R speculometry: to infer either DMSS$_\lambda$ or a combination of DMSS$_\lambda$ and scatterer velocity, depending on the aircraft speed. Further investigations will be carried out to take into account the correct deformation of the Doppler lines on the surface for the search of a scatterer velocity vector. Finally, we emphasize that the flight was not optimized for speculometry: higher and faster flights are needed in the future to consolidate the  DDM inversion  technique  and to test  new higher resolution inversion concepts.
%, as well as, possible detection for sea surface motion Doppler shift signatures as recently revealed using standard Synthetic Aperture Radar Doppler centroid analysis (\markcite {{\it Chapron et al.,} [2003]}).

%% ------------------------------------------------------------------------ %%
%
%  ACKNOWLEDGMENTS
%
%% ------------------------------------------------------------------------ %%

\begin{acknowledgments}
The data analysis and the experimental campaign were respectively carried out under the ESA contracts  3-10120/01/NL/SF (OPPSCAT) and TRP-ETP-137.A. We thank the Institut Cartografic de Catalunya for  flawless flight operations and aircraft GPS/INS kinematic processing. All Starlab authors have contributed significantly; the Starlab author list has been ordered randomly.
\end{acknowledgments}

%% ------------------------------------------------------------------------ %%
%
%  REFERENCE LIST AND TEXT CITATIONS
%
%% ------------------------------------------------------------------------ %%

%% ------------------------------------------------------------------------ %%
%
%  FIGURES
%
%% ------------------------------------------------------------------------ %%

%% ------------------------------------------------------------------------ %%
%
%  END ARTICLE (2/2)
%
%% ------------------------------------------------------------------------ %%
% PLEASE PLACE END ARTICLE COMMAND HERE FOR GALLEY MODE
% FOR DRAFT MODE PLEASE REMEMBER TO (1) COMMENT OUT THIS LINE AND
% (2) PLACE AN END ARTICLE COMMAND AFTER THE BIBLIOGRAPHY INSTEAD
\end{article}


\begin{thebibliography}{10}

\bibitem[{[{\em Armatys et al., 2000}]}]{armatys2000}
Armatys, M.,  A. Komjathy, P. Axelrad, and S. Katzberg. A comparison of GPS
and scatterometer sensing of ocean wind speed and direction. In Proc. IEEE
IGARSS, Honolulu, HA, 2000.

\bibitem[{[{\em Cardellach et al., 2003}]}]{cardellach2003}
Cardellach, E.,  G.~Ruffini, D.~Pino, A.~Rius, A.~Komjathy, and J.~Garrison,
Mediterranean balloon experiment: {GPS} reflection for wind speed
retrieval from the stratosphere.
{\em Rem. Sens. Env.}, 2003.    

\bibitem[{[{\em Chelton et al., 2004}]}]{chelton2004}
Chelton, D.B., M.G. Schlax, M.H. Freilich, and R.F. Milliff,
Satellite Measurements Reveal Persistent Short-Scale Features in Ocean Winds.
{\em Science, 303}, Issue 5660, 978-983, 2004


\bibitem[{[{\em Elfouhaily et al., 1997}]}]{elfouhaily1997}
Elfouhaily, T., B.~Chapron, K.~Katsaros, and D.~Vandemark,  A unified directional spectrum for long and short wind-driven waves, {\em JGR}, 102(15):781--796, 1997.

\bibitem[{[{\em Elfouhaily et al., 2002}]}]{elfouhaily2002}
Elfouhaily, T.,  D.~Thompson, and L.~Linstrom, Delay-{D}oppler analysis of bistatically reflected signals from the  ocean surface: Theory and application.
{\em IEEE TGRS}, 40(3):560--573, 2002.        


\bibitem[{[{\em Garrison, 1998}]}]{garrison1998}
Garrison, J.L., Katzberg, J.L., Effects of sea roughness on bistatically scattered range coded signals from the Global Positioning System, {\em GRL}, vol 25, n. 13, 1998.

\bibitem[{[{\em Garrison, 2002}]}]{garrison2002}
Garrison, J.L., Wind speed measurement using forward scattered {GPS} signals.
{\em IEEE TGRS}, 40:50--65, 2002.

\bibitem[{[{\em Garrison, 2003}]}]{garrison2003}
Garrison, J.L., Anisotropy in Reflected GPS Measurements of Ocean Winds, in {\em Proceedings of the 2003 Workshop on Oceanography with
  GNSS-R}, Starlab, July 2003.  Available at http://starlab.es/gnssr2003/proceedings/. 


\bibitem[{[{\em Gourrion et al., 2002}]}]{gourrion2002}
Gourrion, J., D. Vandemark, S. Bailey, and B. Chapron, Investigation of C-band altimeter cross section dependence on wind speed and sea state, 
Can. J. Rem. Sens., Vol. 28, No. 3, pp. 484-489, 2002.
 
\bibitem[{[{\em Komjathy et al., 2000}]}]{komjathy2000}
Komjathy, A.,  V.~Zavorotny, P.~Axelrad, G.H.~Born, and J.L.~Garrison,
{GPS} signal scattering from sea surface: Wind speed retrieval using
experimental data and theoretical model.
{\em Rem. Sens. Env.}, 73:162--174, 2000.     

\bibitem[{[{\em Ruffini et al., 2000}]}]{ruffini2000}
Ruffini, G., J.L.~Garrison, E.~Cardellach, A.~Rius, M.~Armatys, and D.~Masters.
Inversion of {GPS-R} delay-{D}oppler mapping waveforms for wind retrieval.
{\em In Proc. IEEE IGARSS, Honolulu, HA}, 2000. 


\bibitem[{[{\em Ruffini et al., 2004}]}]{ruffini2004}
Ruffini, G., F.~Soulat, M.~Caparrini, O.~Germain and M.~Martin-Neira , The Eddy Experiment I: {A}ccurate GNSS-R ocean altimetry from low altitude aircraft, {\em to appear in GRL}, 2004.

\bibitem[{[{\em Soulat, 2003}]}]{soulat2003}
Soulat, F., Sea surface remote-sensing with {GNSS} and sunlight reflections, {\em Doctoral Thesis}, Universitat Polit\`{e}cnica de Catalunya/Starlab, 2003.  Available at http://starlab.es/library.html.

\bibitem[{[{\em Soulat, 2004}]}]{soulat2004}
Soulat, F.,  M. Caparrini, O. Germain, P. Lopez-Dekker, M. Taani, G. Ruffini,  
Sea state monitoring using coastal GNSS-R,  
submitted to GRL, http://arxiv.org/abs/physics/0406029. 

\bibitem[{[{\em Watson et al.,  1999}]}]{watson99} 
Watson W. G., et al., NASA/GODDARD Research Activities for the Global Ocean  Carbon Cycle:  A Prospectus for the 21st Century, December 99.

\bibitem[{[{\em Zavorotny et al., 2000}]}]{zavorotny2000}
Zavorotny V., and A.~Voronovich,  Scattering of {GPS} signals from the ocean with wind remote sensing   application,  {\em IEEE TGRS}, 38(2):951--964,   2000.


\bibitem[{[{\em Zuffada et al., 2000}]}]{zuffada2000}
Zuffada, C.,  and T.~Elfouhaily,
Determining wind speed and direction with ocean reflected {GPS} signals,
{\em In Sixth Int. Conf. on Rem. Sens. for Marine and Coastal Environments, Charleston}, 2000.      


\end{thebibliography}
\end{document}